\documentclass[sigconf]{acmart}

\usepackage{subcaption}
\usepackage{balance}
\usepackage{natbib}

\AtBeginDocument{%
  \providecommand\BibTeX{{%
    \normalfont B\kern-0.5em{\scshape i\kern-0.25em b}\kern-0.8em\TeX}}}

\setcopyright{acmcopyright}
\copyrightyear{2021}
\acmYear{2021}
\acmDOI{XX.XXXX/XXXXXXX.XXXXXXX}

\copyrightyear{2022} 
\acmYear{2022} 
\setcopyright{acmcopyright}\acmConference[MSR '22]{19th International Conference on Mining Software Repositories}{May 23--24, 2022}{Pittsburgh, PA, USA} \acmBooktitle{19th International Conference on Mining Software Repositories (MSR '22), May 23--24, 2022, Pittsburgh, PA, USA} \acmPrice{15.00} \acmDOI{10.1145/3524842.3527996} \acmISBN{978-1-4503-9303-4/22/05}

\acmConference[MSR 2022]{MSR '22: Proceedings of the 19th International Conference on Mining Software Repositories}{May 23–24, 2022}{Pittsburgh, PA, USA}

\begin{document}

\title{ApacheJIT: A Large Dataset for Just-In-Time Defect Prediction}

\author{Hossein Keshavarz}
\affiliation{%
  \institution{David R. Cheriton School of Computer Science}
  \institution{University of Waterloo}
  \city{Waterloo}
  \state{Ontario}
  \country{Canada}
}
\email{hossein.keshavarz@uwaterloo.ca}

\author{Meiyappan Nagappan}
\affiliation{%
  \institution{David R. Cheriton School of Computer Science}
  \institution{University of Waterloo}
  \city{Waterloo}
  \state{Ontario}
  \country{Canada}
}
\email{mei.nagappan@uwaterloo.ca}

\renewcommand{\shortauthors}{Keshavarz and Nagappan}

\begin{abstract}
In this paper, we present ApacheJIT, a large dataset for Just-In-Time (JIT) defect prediction. ApacheJIT consists of clean and bug-inducing software changes in 14 popular Apache projects. ApacheJIT has a total of 106,674 commits (28,239 bug-inducing and 78,435 clean commits). Having a large number of commits makes ApacheJIT a suitable dataset for machine learning JIT models, especially deep learning models that require large training sets to effectively generalize the patterns present in the historical data to future data. 
\end{abstract}




\maketitle

\section{Introduction}
\label{sec:intro}
Change-level defect prediction, known as Just-In-Time (JIT) defect prediction, has attracted researchers' attention in recent years \cite{shihab2012-industrial-change-risk,large_scale}. 
JIT defect prediction models are machine learning models relying on historical data. They require a set of past change revisions with each revision being identified whether or not it introduced a bug to the software (bug-inducing). In addition to change revisions and change labels, JIT defect prediction datasets often come with change metrics that have proved to be helpful in analysis and prediction \cite{large_scale,unsupervised-jit,mcintosh&kamei2017}.

Over the past few years, deep learning models found their way to JIT defect prediction \cite{deepjit,cc2vec}. Although deep learning models have demonstrated solid performances in other areas of computing \cite{deep2018,deep2019,deep2021}, DeepJIT \cite{deepjit} and and CC2Vec \cite{cc2vec} do not outperform simple methods like logistic regression  \cite{mcintosh&kamei2017,gema-extrinsic}. This can be attributed to two main reasons. First, there are not many JIT datasets publicly available and most of the existing ones are small; while deep learning models are more effective when the size of the training data is large \cite{datasize2016,datasize2018,datasize2021}. 
Secondly, the number of bug-inducing changes in the lifetime of a software system is often smaller than the number of clean changes. This leads to the class imbalance problem in JIT datasets. Undersampling the majority class makes the dataset even smaller and oversampling the bug-inducing class introduces bias. Deep learning models are more sensitive to both \cite{Johnson2019SurveyOD}.

For example, one of the most widely used datasets for JIT defect prediction is presented by \citet{mcintosh&kamei2017}. Although this dataset consists of carefully curated change revisions in \textit{QT}\footnote{\href{https://www.qt.io/}{https://www.qt.io/}} and \textit{OpenStack}\footnote{\href{https://www.openstack.org/}{https://www.openstack.org/}} projects, it has 25,150 QT changes and 12,374 OpenStack changes, and the ratios of bug-inducing changes to total changes are 8\% and 13\% for QT and OpenStack respectively.


In this work, we present ApacheJIT, a large dataset for JIT defect prediction. ApacheJIT consists of software changes in popular Apache projects. These changes have been selected carefully after applying filtering steps recommended in the literature \cite{szzframework,mcintosh&kamei2017}. ApacheJIT has 106,674 commits (28,239 bug-inducing, 78,435 clean). ApacheJIT is one of the largest available JIT defect prediction datasets and it is suitable for JIT models that require a large number of software changes with many bug-inducing changes.

\section{Related Work}
\label{sec:related}
We found four major JIT datasets in the literature that are large or used in multiple studies.

\citet{large_scale} performed a large-scale study of change-level defect prediction and coined the term \textit{Just-In-Time Quality Assurance}, which evolved into \textit{Just-In-Time Defect Prediction}. They investigated the effectiveness of logistic regression on detecting bug-inducing changes in 6 open-source and 5 commercial software projects. They extracted the changes from CVS and linked the fixing changes to the issues in the issue tracking systems. They used the basic SZZ algorithm \cite{szz} to label bug-inducing changes (except for two open-source projects that did not have issue keys in their change comments and they used Approximate SZZ). The dataset is not publicly available.

\citet{personalized} attempted to separate the prediction for different developers and called this problem \textit{Personalized Defect Prediction}. They built a dataset of Java and C/C++ source codes from 6 open-source projects. The bug-fixing changes of two projects were previously manually found and for the rest of the projects, they applied keyword search. They labeled bug-inducing changes using SZZ without applying any filtering. Although their dataset has been used in \cite{online} and \cite{deep-semantic}, these works are done by the same team and the data is not publicly available. 

\citet{mcintosh&kamei2017} conducted a longitudinal study on JIT bug prediction models to see how the performance of JIT models changes over time. They built a dataset of 37,524 commits from OpenStack (12,374 commits) and QT (25,150 commits). They extended their work in \cite{szzframework} and applied a set of filtering steps on SZZ to remove the false positive bug-inducing changes. The replication package of the study and the datasets are available. This dataset has been widely used to evaluate JIT models. 

\citet{mislabeled} investigated the impact of mislabeled changes labeled by four SZZ variants (Basic SZZ, AG-SZZ, MA-SZZ, RA-SZZ). They claim that RA-SZZ generates the cleanest labels and used this variant as the baseline. They did not include \citet{mcintosh&kamei2017}'s variant in the study because it does not address false negatives (due to code indentation). They built a dataset of 10 Apache projects with 126,526 commits. RA-SZZ (the baseline) identifies 13,078 bug-inducing commits in this data. Although the authors have made the data available, the link between revision IDs and bug-inducing labels is missing. To the best of our knowledge, this data is not used in any JIT model evaluation.

In this work, we adopt the \citet{mcintosh&kamei2017}'s approach to identify bug-inducing commits because the dataset has been widely used to evaluate JIT models \cite{deepjit,cc2vec,pornprasit2021jitline,jitbias}.

\vspace{-0.23in}
\section{ApacheJIT Dataset \& Usage}
\label{sec:dataset}
The current work presents the \textit{ApacheJIT} dataset. ApacheJIT is one of the largest available datasets for JIT defect prediction. This dataset is a collection of carefully selected and filtered software changes in a set of popular Apache projects. ApacheJIT includes 106,674 software revisions from 2003 to 2019. These change revisions are derived from the issue reports from January 1, 2010, to December 31, 2019. 28,239 of these revisions are labeled as bug-inducing through the process explained in Section \ref{sec:construction}. ApacheJIT is suitable for defect prediction models that require a large set of historical data to learn prediction models. 

In particular, ApacheJIT can be used to train deep learning models that require large datasets for effectively capturing the patterns in the historical data and using them to accurately predict future observations. Currently, the performances of deep learning models on JIT defect prediction datasets are not as promising as their performances in other areas of computing. One reason is that available JIT defect prediction datasets do not contain many samples and consequently, the number of bug-inducing changes models see during the training is very small. 


In addition to identifying whether or not each change revision has introduced bugs into systems, the data presented in this work also includes some of the change metrics that are commonly used for JIT defect prediction. The following is the list of these metrics (columns of the datasets): 

\textit{change date, \# of lines added, \# lines deleted, \# files touched, \# directories touched, \# of subsystems touched, change entropy, \# of distinct developers touched files, the average time from last change, \# of unique changes in files, change author experience, change author recent experience, change author subsystem experience}.

The explanation of each metric is presented in \citet{large_scale}. We used the same approach to obtain the change metrics in this work. 
Table \ref{table:issue-bug} shows the statistics of ApacheJIT. The ApacheJIT dataset and the related scripts are publicly available\footnote{\href{https://doi.org/10.5281/zenodo.5907001}{https://doi.org/10.5281/zenodo.5907001}}.


\section{Data Construction}
\label{sec:construction}
The major part of constructing the ApacheJIT dataset is finding bug-inducing commits. This part was done based on the SZZ algorithm \cite{szz}. The SZZ algorithm has been widely used to detect bug-inducing commits, and in this work, we used it with some modifications. The SZZ algorithm starts with collecting the issue reports that are marked as fixed. Then these fixed issue reports are linked to their corresponding fixing commits. And finally, from the lines changed in the fixing commits, potential bug-inducing commits are detected.

Initially, we selected 15 popular Apache projects that had many bug reports (we used the data of 14 project in the end). Our measure of popularity in this selection was the number of stars each project has on GitHub. Table \ref{table:projects} shows the selected projects.


\begin{table}[htbp]
\caption{Selected Apache projects in this study}
\begin{center}
\begin{tabular}{|c|c|c|c|c|}
\hline
ActiveMQ & Camel & Cassandra & Flink & Spark  \\ \hline
Zeppelin & Groovy & Hadoop HDFS & HBase & Hive \\ \hline
Zookeeper & Ignite & Hadoop MapReduce & Mesos* & Kafka  \\ \hline
\end{tabular}
  {\raggedright \small * removed after collecting fixing commits (Section \ref{subsec:fixing}) \par}


\label{table:projects}
\end{center}
\end{table}

\vspace{-0.2in}
\subsection{Bug Report Collection}
\label{subsec:issues}

As explained above, SZZ starts with collecting issue reports.
All the Apache projects we selected keep their issue reports on Apache's JIRA Issue Tracker\footnote{\href{https://issues.apache.org/jira/}{https://issues.apache.org/jira/}}. On JIRA, after selecting the aforementioned projects, we applied further filtering. First, we narrowed down our study focus to the issue reported from January 1, 2010, to December 31, 2019. Next, we filtered out the issues that were not identified as \textit{bug}s. And finally, we picked the issues that are fixed. On JIRA, these issues are the issues with Status set to \textit{Closed} or \textit{Resolved} and with Resolution set to \textit{Fixed}.
Finally, after applying the filtering steps mentioned above, we had 56,929 bug reports. Table \ref{table:issue-bug} shows the number of bug reports (issues) in each project.

\subsection{Fixing Commit Collection}
\label{subsec:fixing}

After obtaining the issue reports that have been fixed, we looked for the commits that fixed these issue reports in the version control system (VCS). All these projects use Git as their VCS. We followed the approach in \cite{large_scale} and \cite{mcintosh&kamei2017}. Each issue is identified uniquely with an issue key on JIRA. With the help of this identifier, for each project, we searched through all the commits on the main branch of the project Git repository and looked for commits whose commit messages indicate the change is fixing one of the issue keys we collected. This approach works because conventionally, developers add the issue key of the bugs they fix to the commit message. 

In previous work, the search for these commits is done manually by looking for keywords in the result of the \textit{git log} command \cite{szz,large_scale,mcintosh&kamei2017}. In this work, however, we utilized GitHub search. This was feasible because all the projects we selected are stored on GitHub. The reason we preferred GitHub search to manual pattern matching was that the GitHub search engine returns the best match if it exists. This is especially useful when a commit message of a fixing commit does not include the issue key in the expected format. We compared the result of the GitHub search with the result of string matching on \textit{git log} outputs and found out that overall, the commits returned by GitHub search are more relevant. To search on GitHub, we used the GitHub REST API\footnote{\href{https://docs.github.com/en/rest}{https://docs.github.com/en/rest}}.

In this process, if there is no commit with a commit message that contains an issue key, GitHub returns no commit. If there are commits with commit messages containing an issue key, GitHub returns one or several commits. We could identify the following reasons for the latter case:
\begin{enumerate}
    \item Developers make several attempts to fix an issue before closing the bug report because the first attempts are \textit{not enough}.
    \item Developers make several attempts to fix an issue before closing the bug report because the first attempts are \textit{incorrect}.
\end{enumerate}

The scenarios mentioned above make finding true fixing commits challenging in both pattern matching and GitHub search approaches. On one hand, one may decide to include all the commits returned for one issue key because they are all related. On the other hand, one may only consider the most relevant of the multiple commits as the fixing commit. In this work, we chose the latter direction and picked the latest commit as the fixing commit. Our justification was that by picking all the aforementioned commits, later SZZ will consider many clean commits as bug-inducing (case 2 above). Therefore, the ultimate dataset will have high false positive bug-inducing commits. 

Among all the commits returned for one bug report, we found the latest one to be the most relevant. By picking the latest commit as the fix commit, we are almost certain that the commit we have picked is truly a fixing commit and can be used to find bug-inducing commits based on SZZ. This approach, however, leads to missing some bug-inducing commits (case 1 above), and consequently, our ultimate dataset will have higher false negative commits (bug-inducing commits that are labeled as clean). Essentially, this is a trade-off between more false positives and more false negatives, and in this study, we chose the latter.

After collecting fixing commits as described above for all 15 projects we noticed that the commit messages in Apache Mesos do not comply with the conventional format and even GitHub search was not able to find fixing commits. Therefore, we eliminated all Apache Mesos data and continued with the remaining 14 projects. 

\begin{table}[htbp]
\caption{The number of collected issues and the statistics of ApacheJIT. Percentages under the Bug-inducing column indicate the ratio of bug-inducing commits to total commits.}
\begin{center}
\begin{tabular}{c|c|c|c|c}

\textbf{Project} & \textbf{Issues} & \textbf{Bug-inducing} & \textbf{Clean} & \textbf{Total} \\
\hline
ActiveMQ & 1,967 & 1,404 (23\%) & 4,722 & 6,126 \\ \hline
Camel & 3,276 & 3,078 (14\%) & 19,622 & 22,700 \\ \hline
Cassandra & 5,358 & 3,117 (38\%) & 5,042 & 8,159 \\ \hline
Flink & 4,166 & 2,811 (24\%) & 8,880 & 11,691 \\ \hline
Groovy & 2,549 & 1,614 (20\%) & 6,445 & 8,059 \\ \hline
HDFS & 3,672 & 2,222 (21\%) & 8,137 & 10,359 \\ \hline
HBase & 7,085 & 3,782 (43\%) & 4,948 & 8,730 \\ \hline
Hive & 7,931 & 4,223 (61\%) & 2,619 & 6,842 \\ \hline
Ignite & 3,256 & 2,439 (20\%) & 9,597 & 12,036 \\ \hline
MapReduce & 2,080 & 838 (15\%) & 4,995 & 5,833 \\ \hline
Mesos & 2,955 & - & - & - \\ \hline
Kafka & 3,038 & 1,115 (46\%) & 1,269 & 2,384 \\ \hline
Spark & 7,648 & 632 (43\%) & 833 & 1,465 \\ \hline
Zeppelin & 1,089 & 622 (42\%) & 829 & 1,451 \\ \hline
Zookeeper & 859 & 342 (40\%) & 497 & 839 \\ \hline
\textbf{Total} & \textbf{56,929} & \textbf{28,239 (26\%)} & \textbf{78,435} & \textbf{106,674}
\end{tabular}
\label{table:issue-bug}
\end{center}
\end{table}

\subsection{Finding Bug-inducing Commits}
\label{subsec:buggy}

At the end of the previous step, we linked 44,202 commits in the 14 projects to issue keys. These commits represent the fixing commits for the collected issues. The next step is to use these fixing commits to find bug-inducing commits.

\subsubsection{Git Annotate}

In this step, each fixing commit is traced using the \textit{git annotate} command. This command annotates all lines of a given file showing the last revisions that touched each line. For each fixing commit, we ran \textit{git annotate} on the files modified in the commit and obtained the last revisions that touched the deleted lines before the fixing commit. 
To implement this process, we used the SZZ tool in the PyDriller framework\footnote{\href{https://pydriller.readthedocs.io/}{https://pydriller.readthedocs.io/}} \cite{pydriller}. The SZZ tool in PyDriller gets a commit and returns the commits that last touched the deleted lines of the files modified in the given commit. 

\subsubsection{Filtering} 
\label{subsec:filter}
The basic version of SZZ has limitations. The SZZ algorithm tends to label many clean commits as bug-inducing. Accordingly, we performed some heuristics to reduce false positives. We followed the filtering discussed in \cite{szzframework} and \cite{mcintosh&kamei2017} to filter out linked bug-inducing commits that are likely to be clean.

\begin{enumerate}
    \item We made fixing commit - bug-inducing commit pairs and associated each with the issue key corresponding to the fixing commit. We removed the pairs where the bug-inducing commit date was after the issue report date (the date the issue was created on JIRA). Note that we removed the pairs and not the bug-inducing commits or the fixing commits separately as they may show up in other pairs and end up as valid bug-inducing and valid fixing commits respectively. This step filtered 5,048 bug-inducing commit candidates.

    \item At the end of the \textit{git annotate} process each fixing commit may be linked to several bug-inducing commits (a fixing commit may fix several bugs). We call the number of bug-inducing commits each fixing commit is linked to \textit{fixcount}. \citet{szzframework} and \citet{mcintosh&kamei2017} filter out fixing commits whose fixcounts are more than a threshold. They refer to these commits as \textit{suspicious fixing commits}. In their works, the threshold is set to upper Median Absolute Deviation (MAD) of fixcounts.
    
    \vspace{-0.1in}
    \begin{equation}
        upper MAD = M + median(|M - X_i|),
    \end{equation}
    
    where $M$ is the median of fixcounts and $X_i$ is the fixcount of commit $i$.
    
    In the present work, however, the upper MAD was too small, and choosing it as the threshold would filter too many fixing commits. As an alternative, we chose the sum of the mean and the standard deviation of fixcounts as our threshold. Note that again, this is a trade-off between high false positive and high false negative. Filtering out too many fixing commits will cause more bug-inducing commits to be labeled as clean commits in later steps. This step filtered 12,165 commits.
    
    \item Similarly, we can define \textit{bugcount} as the number of fixing commits each bug-inducing commit is linked to. This means that one commit has introduced multiple bugs into the system (multiple bug reports) and each has been fixed by a fixing commit. Again, to filter out the suspicious bug-inducing commits, we set a threshold of $mean + std$ of bugcounts. This step removed 1,257 bug-inducing commit candidates.
    
    \item Following \citet{mcintosh&kamei2017}, we removed large commits. By large commits, here, we mean the commits that modify more than 100 files or have more than 10,000 lines of changed code. Lines of changed code is the total number of lines that were removed or added through the commit. This step removed 890 bug-inducing commit candidates.
    
    \item In this work, we focused on Java programming language and built a uniform Java dataset. Language constraint makes it feasible to do static analysis on the source codes (our static analysis is explained in the next part). Among the selected projects, Java is the language in which most of the repository files are written. Therefore, we picked Java and filtered out the commits that do not modify any Java source code. 10,251 commits were filtered.
    
    \item To avoid trivial changes, we performed a static analysis on the abstract syntax trees (ASTs) of the changed source code. In this process, we compared the AST of each Java program that was changed in a commit before and after the change. If there was at least one node in either of the two ASTs without a match in the other AST, we marked the change as non-trivial and kept the Java source code; otherwise, the change is trivial. We filtered the commit if all the Java source codes in it were changed trivially. To conduct this analysis, we used  
    GumTreeDiff \cite{gumtree}. GumTreeDiff is a tool that finds the differences between two source codes written in the same language using their ASTs. 
    Examples of trivial changes are changes that modify comments, white spaces, and string or numeric literals. This step removed 274 commits.
    
\end{enumerate}

Before applying the filtering steps, there were 58,124 bug-inducing commit candidates. 29,885 commits were filtered and the remaining 28,239 commits were labeled as bug-inducing commits. Table \ref{table:issue-bug} shows the number of bug-inducing commits in each project. It is worth mentioning that the number of bug-inducing commits in Apache Spark is significantly low with respect to the number of issue reports in this project. The reason is that Java is not the major programming language in Apache Spark.

\subsection{Finding Clean Commits}
\label{subsec:clean}

In addition to the commits that introduce bugs into the system, we also would like to know what changes are safe. We call these changes \textit{clean} commits. 
Usually in projects, the clean commits outnumber bug-inducing commits because most changes are reviewed before they are integrated into the system. 

We collected all the commits in the selected project repositories from the date of the earliest bug-inducing commit (Sep 11, 2003) to the date of the latest one (Dec 26, 2019). 
We removed the commits we already had labeled as bug-inducing and the fixing commits that had at least one corresponding bug-inducing commit to avoid bias (they change the same set of lines as their corresponding bug-inducing commits). The number of remaining commits was 149,962. We also applied filtering steps (4) and (5) in Section \ref{subsec:filter} to make the filtering process similar to bug-inducing commits (steps (1)-(3) are not applicable). Finally, we had 78,435 clean commits. 



\subsection{Commit Metrics}

In the end, we also added the set of common change metrics defect prediction datasets have. The metrics we collected were the same as the ones discussed in \cite{large_scale}. We followed the same steps.

\section{Limitations}
\label{sec:limitation}
Like previous work, we used the SZZ algorithm to find bug-inducing changes in this work. Prior studies have discussed the limitations of SZZ \cite{szzframework,ra-szz,ra-szz-revisited}. However, SZZ is still the most commonly used algorithm for finding bug-inducing commits. We followed the filtering steps that are shown to be effective in removing safe changes that are identified as bug-inducing by SZZ \cite{szzframework,mcintosh&kamei2017}. Moreover, in this work we utilized GumTreeDiff \cite{gumtree} to remove trivial changes. GumTreeDiff is a powerful tool that uses abstract syntax trees (ASTs) of programs to find the structural differences between two codes; however, it does not support all programming languages\footnote{\href{https://github.com/GumTreeDiff/gumtree/wiki/Languages}{https://github.com/GumTreeDiff/gumtree/wiki/Languages}}.

Finally, in this work, the focus is on intrinsic bugs. \citet{gema-extrinsic} study another class of software bugs, called extrinsic bugs, that are not easily detected. Extrinsic bugs directly affect the performance of defect prediction models. As discussed in that paper, the SZZ algorithm is unable to identify extrinsic bugs.

\section{Conclusion}
\label{sec:conclusion}
Among different forms of software defect prediction, Just-In-Time (JIT) defect prediction has gained popularity in recent years. Models attempting to predict whether a change is likely to introduce bugs into the system are machine learning models trained on historical data. Currently, most of the available datasets are small and it is difficult for JIT models to effectively learn from historical data and generalize to future data. In this work, we present ApacheJIT, a large JIT defect prediction dataset with 106,674 software changes. ApacheJIT has 28,239 bug-inducing changes. The dataset is available here: \href{https://doi.org/10.5281/zenodo.5907001}{https://doi.org/10.5281/zenodo.5907001}

\begin{acks}
The authors would like to thank Dr. Gema Rodríguez-Pérez, Dr. Shane McIntosh, and Dr. Yasutaka Kamei for their invaluable help in collecting this dataset. The authors also acknowledge that our work takes place on the traditional territory of the Neutral, Anishinaabeg and Haudenosaunee peoples.
\end{acks}

\newpage
\balance
\bibliographystyle{ACM-Reference-Format}
\bibliography{main}

\end{document}